\documentclass[spmi]{academic}
\usepackage{epsfig}

\begin{document}

\shortauthor{Mortensen and Bastian}

\shorttitle{\it Side-gate modulation of critical current ...}

\title{ Side-gate modulation of critical current in mesoscopic
  Josephson junction }

\author[1]{Niels Asger Mortensen}

\author[2,3]{Georg Bastian}

\address[1]{Mikroelektronik Centret, Technical University of Denmark,
  \O rsteds Plads, Bld. 345 east, DK-2800 Kgs. Lyngby, Denmark}

\address[2]{Physikalisch-Technische Bundesanstalt, Bundesallee 100,
  D-38116 Braunschweig, Germany}

\address[3]{NTT Basic Research Labs, 3-1 Morinosato Wakamiya, Atsugi,
  Kanagawa, 243-0198 Japan}


\maketitle

\begin{abstract}
  We study the normal state conductance and the Josephson current in
  a superconductor-2DEG-superconductor structure where the
  size/shape of the 2DEG-region can be modified by an additional
  side-gate electrode. The considered transport properties follow from
  the retarded Green function which we compute by employing a
  tight-binding-like representation of the Hamiltonian in the 2DEG
  region. Our model studies offer a qualitative demonstration of the
  recently observed effects caused by side-gate modulation.
\end{abstract}

\section{Introduction}

Devices with semiconducting materials as barriers in Josephson weak
links offer the possibility to modulate the junction properties in
order to investigate both fundamental transport mechanisms as well as
for use in applications of superconducting devices. Several methods
for tunable Josephson junctions have been studied both theoretically
and experimentally. Tunable effects have been observed in {\it e.g.}
Josephson-field-effect-transistors \cite{clark1980}, non-equilibrium
junctions \cite{morpurgo1998,kutchinsky1999}, optically modulated weak
links \cite{bastian1999}, and recently in side-gate modulated junctions
\cite{bastian2000}.

Following Ref. \cite{bastian2000} we consider
superconductor-2DEG-superconductor structures where the indirect
ballistic transport between the non-opposite superconducting contacts
can be controlled by a voltage applied to a side-gate (see Fig.
\ref{FIG2}). The geometry has many similarities with the T-stub wave
guide geometry \cite{sols1989,aihara1993} but here the two interfaces
to the superconductors are non-opposite. The aim is to demonstrate the
principle behind side-gate modulation of the critical current and also
to investigate the local density-of-states in the 2DEG in relation to
the discussion of ``non-local'' modes \cite{heida1998,bastian1998}.

In this work we focus on the normal state conductance and the
Josephson current in the case where electron and hole propagation in
the normal region is phase-coherent. For convenience we consider
the zero temperature limit. The normal state conductance is then given
by the Landauer formula \cite{landauer1957,landauer1970,fisher1981}
which at $T=0\,{\rm K}$ has the particularly simple form

\begin{equation}\label{GN}
G_{\rm N} = \frac{2e^2}{h} {\rm Tr}\,tt^\dagger 
= \frac{2e^2}{h} \sum_{n=1}^N T_n,
\end{equation}
where $2e^2/h$ is the quantum unit of conductance and $T_n$
($n=1,2,\ldots N$) are the transmission eigenvalues of $tt^\dagger$
which is an $N\times N$ matrix, $N$ being the number of propagating
modes at the Fermi level. Beenakker and van Houten
\cite{beenakker1991a} considered the Josephson current through a
superconducting quantum point contact and found that the critical
current was quantized in units of $e\Delta/\hbar$, $\Delta$ being the
energy gap of the superconductor. Subsequently it was shown by
Beenakker \cite{beenakker1991b} that just like the normal state
conductance, also the Josephson current between two superconducting
leads can also be expressed in terms of $T_n$ of the normal region
which couples the two superconductors. At $T=0\,{\rm K}$ the Josephson
current can be written as \cite{beenakker1991b}

\begin{eqnarray}\label{IJ}
I_{\rm J}(\phi)&=&\frac{e\Delta}{2\hbar}\sin\phi {\rm Tr}\, 
tt^\dagger\left[\hat{1}-tt^\dagger \sin^2(\phi/2)\right]^{-1/2}\nonumber\\
&=&\frac{e\Delta}{2\hbar}\sin\phi \sum_{n=1}^N 
T_n\left[1-T_n\sin^2(\phi/2)\right]^{-1/2},
\end{eqnarray}
where $\phi$ is the phase difference between the two superconductors.
Eq. (\ref{IJ}) is valid in the short-junction limit where the
dimension of the normal region is short compared to the
superconducting coherence length $\xi$ which in the ballistic regime
is $\xi = \hbar v_{\rm F}/\pi\Delta$, $v_{\rm F}$ being the Fermi
velocity.

We note that in Eqs.  (\ref{GN}) and (\ref{IJ}) all transverse
degrees-of-freedom are incorporated in the transmission amplitude
matrix $t$ and that the angle-dependence of the Andreev scattering
discussed in Ref.  \cite{mortensen1999a} follows directly from the
angle (mode) dependence of $t$. Scattering due to non-matching Fermi
velocities or Fermi momenta can be included in several ways: {\it i)}
by calculating the transmission from the composite scattering matrix
$S=S_{{\rm I}_1}\otimes S_{\rm 2DEG}\otimes S_{{\rm I}_2}$ where the
two interfaces (${\rm I}_1$ and ${\rm I}_2$) are described by the
scattering matrices $S_{\rm I_1}$ and $S_{\rm I_2}$ as in Ref.
\cite{mortensen1999b}, {\it ii)} by adding an effective potential to
the 2DEG at the interfaces, or {\it iii)} by explicitly taking the
different Fermi velocities and Fermi momenta into account.  In this
work we will however for simplicity assume matching Fermi properties
of the superconductor and the 2DEG.

\begin{figure}
\begin{center}
 \epsfig{file=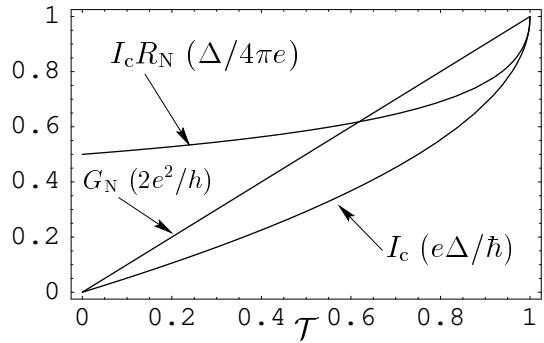, width=0.5\columnwidth,clip}
\end{center}
\caption{Plot of $G_{\rm N}$, $I_{\rm c}$, and $I_{\rm c}R_{\rm N}$ as a function of transmission probability ${\cal T}=|t|^2$ for a system with single-mode leads.}
\label{FIG1}
\end{figure}

\begin{figure}
\begin{center}
\epsfig{file=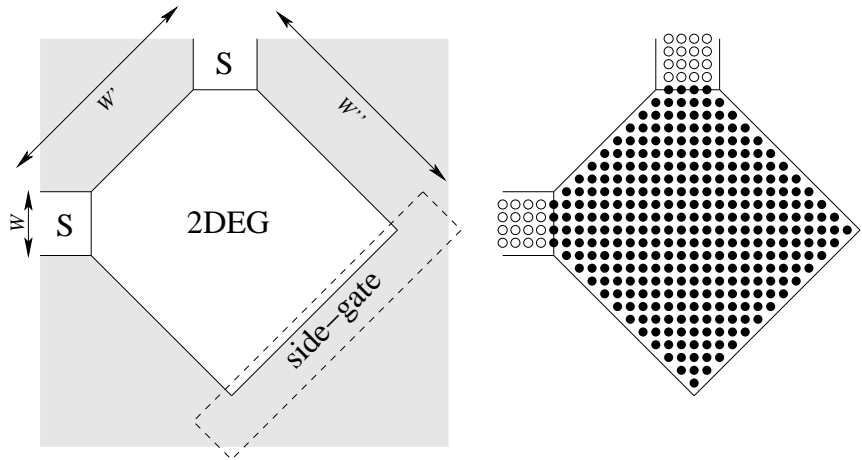, width=0.8\columnwidth,clip}
\end{center}
\caption{Indirect S-2DEG-S structure with a side-gate (left) and
the corresponding lattice model (right).} \label{FIG2}
\end{figure}

To illustrate how the transport properties depend on the transmission
let us consider the case of single-mode leads. Fig.  \ref{FIG1} shows
how the normal state conductance $G_{\rm N}$, the critical current
$I_{\rm c}=\max I_{\rm J}(\phi)$, and the product $I_{\rm c} R_{\rm
  N}=I_{\rm c}/G_{\rm N}$ depend on the transmission probability
${\cal T}=|t|^2$. Compared to the off-resonance regime with ${\cal
  T}\sim 0$ (the Ambegaokar--Baratoff regime \cite{ambegaokar1963})
the $I_{\rm c} R_{\rm N}$ product is enhanced by a factor-of-two when
the 2DEG region is tuned to resonance (${\cal T}\sim 1$). This
factor-of-two enhancement can be considered as a signature of Andreev
mediated transport at resonance whereas the transport is
tunneling-like at off-resonance.

Equations (\ref{GN}) and (\ref{IJ}) form the basis for our
calculations of the transport properties and since they only depend on
the normal state transmission properties of the 2DEG region we can
employ standard methods for quantum transport in semiconductor
structures. The transmission can conveniently be calculated from the
retarded Green function ${\cal G}_\varepsilon^r({\bf r}_1,{\bf r}_2)$
\cite{fisher1981} and also the local density-of-states follows from
${\cal G}_\varepsilon^r({\bf r}_1,{\bf r}_2)$. We calculate ${\cal
  G}_\varepsilon^r({\bf r}_1,{\bf r}_2)$ by applying a finite
differences method to the Hamiltonian of the 2DEG region.

The paper is organized as follows: Section II introduces the finite
differences method and formulates Eqs. (\ref{GN}) and (\ref{IJ}) in
terms of the retarded Green function, section III presents our
geometry and model, section IV contains our results, and in section V
discussions and conclusions are given.

\section{Finite differences method}

We discretize the continuous Hamiltonian by writing the Laplacian as
finite differences \cite{datta}. This gives rise to a
tight-binding-like representation

\begin{eqnarray}
\left\{H_c\right\}_{ij}&=& z\times \gamma +U({\bf r}_i)\hspace{10mm} i=j, \\
&=&-\gamma \hspace{26mm} i,j \,{\rm NN},\nonumber\\
&=&0\hspace{28mm}{\rm otherwise}\nonumber,
\end{eqnarray}
where $z=4$ is the number of nearest neighbors (NN) and
$\gamma=\hbar^2/2ma^2$ corresponds to a hopping matrix element, $a$ being
the lattice spacing. Comparing the energy dispersion to the parabolic
dispersion of the continuous problem shows that the finite differences
method is a very accurate description for energies $\varepsilon<\gamma$
\cite{datta,ferry}. Making the grid finer ($a$ smaller) improves the
accuracy and/or allows for a treatment of higher energies.

With the above definition of the Hamiltonian we label the lattice
points by a number from $1$ to $M$, $M$ being the number of lattice
points. The retarded Green function can then be written as an $M\times
M$ matrix \cite{datta,ferry}

\begin{equation}\label{Gr}
{\mathcal G}^{\rm r}(\varepsilon) = 
\left[\varepsilon \hat{1}-H_c-\Sigma^{\rm r}(\varepsilon)\right]^{-1},
\end{equation}
where the element $\left\{ {\mathcal G}^{\rm r} (\varepsilon)
\right\}_{ij}$ corresponds to ${\cal G}_\varepsilon^r({\bf r}_i,{\bf
  r}_j)$. Here, $H_c$ is the tight-binding Hamiltonian of the
conductor (describing the conductor as a closed system) and
$\Sigma^{\rm r}=\Sigma_1^{\rm r}+\Sigma_2^{\rm r}$ is the self-energy
describing the coupling to leads $1$ and $2$. For the self-energies
the only non-zero elements are those where the leads couple to the
2DEG (scattering region). For large $M$ it can be useful to employ a
recursive method for calculating ${\mathcal G}^{\rm r}(\varepsilon)$
but even for $M\sim 1000$ the direct matrix inversion in Eq.
(\ref{Gr}) can be done without too much numerical effort.

The transmission amplitude matrix $t$ can now be calculated from the
Fisher--Lee relation \cite{fisher1981} which gives
\cite{fisher1981,datta,ferry,brandbyge1999}

\begin{equation}
t(\varepsilon)=\left[\Gamma_1(\varepsilon)\right]^{1/2}
{\cal G}^{\rm r}(\varepsilon) \left[\Gamma_2(\varepsilon)\right]^{1/2},
\end{equation}
where $\Gamma_j(\varepsilon)=i\left(\Sigma_j^{\rm
    r}(\varepsilon)-[\Sigma_j^{\rm r}(\varepsilon)]^\dagger\right)$.
Substituting into Eqs.  (\ref{GN}) and (\ref{IJ}) gives

\begin{equation}
G_{\rm N} = \frac{2e^2}{h} 
{\rm Tr}\, \Gamma_1(\varepsilon) {\cal G}^{\rm r}(\varepsilon) 
\Gamma_2(\varepsilon)[{\cal G}^{\rm r}(\varepsilon)]^\dagger,
\end{equation}
and

\begin{eqnarray}
I_{\rm J}(\phi)&=&\frac{e\Delta}{2\hbar}\sin\phi\, {\rm Tr}\, 
\Gamma_1(\varepsilon) {\cal G}^{\rm r}(\varepsilon) 
\Gamma_2(\varepsilon)[{\cal G}^{\rm r}(\varepsilon)]^\dagger\\
&&\times \left[\hat{1}-\Gamma_1(\varepsilon) {\cal G}^{\rm r}(\varepsilon) 
\Gamma_2(\varepsilon)[{\cal G}^{\rm r}(\varepsilon)]^\dagger 
\sin^2(\phi/2)\right]^{-1/2},\nonumber
\end{eqnarray}
where we have used the cyclic invariance of the trace. Once the
retarded Green has been obtained (by a single matrix inversion) these
relations directly provide the essential transport properties. The
local density-of-states (in the normal state) can also be calculated
directly from the retarded Green function (see {\it e.g.} \cite{datta})

\begin{equation}
\rho({\bf r}_j,\varepsilon)=-\frac{1}{\pi} {\rm Im} \left\{ {\cal G}^{\rm r}(\varepsilon)\right\}_{jj},
\end{equation}
and this function is useful in obtaining insight into the spatial
variations of the states and revealing the nature of the conduction
through the sample.

\section{Geometry and model}

We consider the geometry shown in Fig. \ref{FIG2}. The 2DEG is
confined also in the direction parallel to the side-gate so that the
system acts as a cavity or quantum dot coupled to two superconductors
and an additional side-gate. We note that our geometry is slightly
different from that of Ref.  \cite{bastian2000} where there is no
confinement of the 2DEG in the direction parallel to the side-gate.
This means that here the side-gate is used in tuning the cavity to
resonance whereas in Ref. \cite{bastian2000} it rather acts as a
'classical' mirror which can be used in focusing the incident wave
from one lead onto the other lead.

We treat the transmission problem fully quantum mechanically by means
of the presented finite differences method and the lattice version of
the sample shown in Fig. \ref{FIG2}. In this case $a=W/(M_{\rm
  L}+1)$ with $M_{\rm L}=4$ sites in the transverse direction of the
leads of width $W$. In the continuous description of the leads $N={\rm
  Int}(k_{\rm F}W/\pi)$ gives the number of propagating modes at the
Fermi level, ${\rm Int}(x)$ being the integer part of $x$. The
dimensions are $W'=13\sqrt{2}a\sim 3.7 \times W$ and
$W''=(13+1/2)\sqrt{2}a\sim 3.8\times W$. We choose $W$ and the Fermi
level such that $N=1$. The lattice model then gives a reasonable
description of the continuous problem -- the threshold energy $E_1$ of
the first mode deviates only by $\sim 3\%$ from the continuous result.

The side-probe is assumed to act as a gate and only affect the
potential $U$ of the 2DEG through an electro-static coupling. Here, we
note that fluctuations in the gate potential may lead to dephasing of
the electron and hole propagation in the 2DEG (see {\it e.g.} Ref.
\cite{buttiker2000}) and the same can also be the case if the probe
acts as a voltage probe (see {\it e.g.} Ref.  \cite{mortensen2000}).
Here, we use the model in Ref.  \cite{glazman1992} [Eq. (19)] to
account for the potential modification in the 2DEG due to the
side-gate. The distance $\zeta$ from the side-gate to the 2DEG edge
position is then given by

\begin{equation}
\zeta =\frac{\epsilon V_{\rm g}}{4\pi^2 n_0|e|}
=\frac{\epsilon V_{\rm g}}{2\pi k_{\rm F}^2 |e|},
\end{equation}
where $V_{\rm g}$ is the side-gate voltage,
$\epsilon=\epsilon_r\epsilon_0$ is the dielectric constant of the
semiconductor ($\epsilon_r\sim 14$ for GaAs), and $n_0=k_{\rm
  F}^2/2\pi$ the 2DEG density. With the lattice shown in Fig.
\ref{FIG2} the distance $\zeta$ can be changed in steps of
$\delta\zeta = a/\sqrt{2}$ corresponding to $\delta V_{\rm
  g}=\sqrt{2}\pi k_{\rm F}^2 a |e|/\epsilon$.

\section{Results}

\begin{figure}
\begin{center}
\epsfig{file=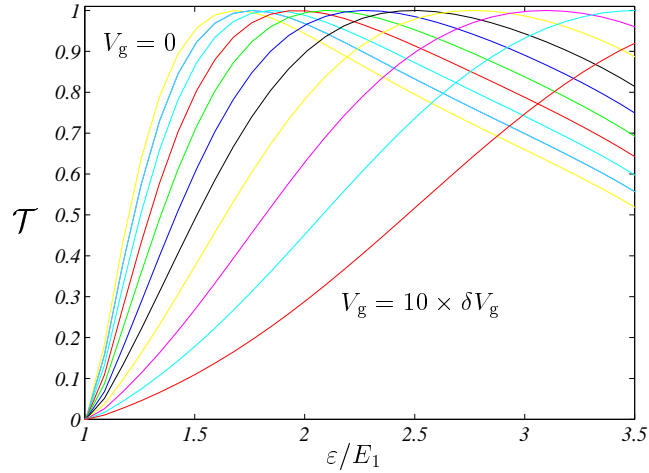, width=0.6\columnwidth,clip}
\end{center}
\caption{The transmission of the lowest mode ${\cal T}=|t|^2$ as a function of the energy $\varepsilon$ (normalized by the threshold energy $E_1$) for gate voltages $V_{\rm g}=j\times \delta V_{\rm g}$ with $j=0,1,2,\ldots 10$.}
\label{FIG3}
\end{figure}

In Fig. \ref{FIG3} we show the transmission as a function of the
energy for different values of the gate voltage. Since the interfaces
between the leads and the 2DEG are fully transparent the transmission
shows a broad resonance behavior. By changing the gate voltage the
cavity is 'squeezed' and as expected the resonance shifts towards higher
energies. For a given Fermi level the side-gate can thus also be used
in tuning the transmission to resonance.

\begin{figure}
\begin{center}
\epsfig{file=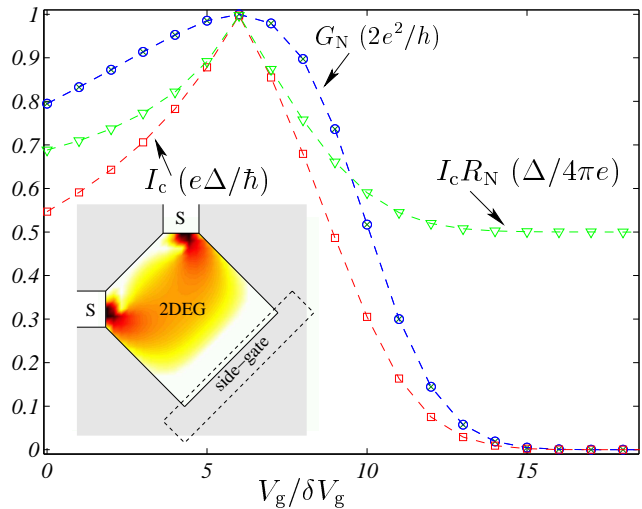, width=0.6\columnwidth,clip}
\end{center}
\caption{The normal state conductance $G_{\rm N}$,
  the critical current $I_{\rm c}$, and the $I_{\rm c}R_{\rm N}$
  product as a function of gate voltage $V_{\rm g}$ for an energy
  $\varepsilon=2.5\times E_1$. The inset shows the local density of
  states $\rho({\bf r},\varepsilon)$ (in the normal state) at gate
  voltage $V_{\rm g}=6\times \delta V_{\rm g}$ corresponding to the
  resonance condition. Dark regions indicate a high density of states  ({\it e.g.} near the leads) and bright regions a lower ({\it e.g.} near the
  side-gate).}
\label{FIG4}
\end{figure}

In Fig. \ref{FIG4} we show the normal state conductance $G_{\rm N}$,
the critical current $I_{\rm c}$, and the $I_{\rm c}R_{\rm N}$ product
as a function of gate voltage for an energy $\varepsilon=2.5\times
E_1$. By increasing the gate voltage the cavity is tuned to resonance
at $V_{\rm g}\sim 6 \times \delta V_{\rm g}$ where the normal state
conductance equals the quantum unit of conductance $2e^2/h$ and the
critical current equals the quantum unit of critical current
$e\Delta/\hbar$. The $I_{\rm c}R_{\rm N}$ product changes by a
factor-of-two from the Ambegaokar--Baratoff value $\Delta/8\pi e$ to
its quantum unit $\Delta/4\pi e$ at resonance.

In the inset of Fig. \ref{FIG4} we show the local density of states
$\rho({\bf r},\varepsilon)$ (in the normal state) at the energy
$\varepsilon=2.5\times E_1$ for a gate voltage corresponding to the
resonance condition. The transparent interfaces give rise to a local
density of states forming a relatively smooth and connected 'path'
between the two leads. The plot also clearly shows the wave nature of
the electron and hole propagation. Thus a more simple semi-classical
trajectory model with point-like particles and a side-gate acting as a
classical mirror would be inadequate for the present situation. This
is often the case and as studied in {\it e.g.} Ref. \cite{jauho1999}
complicated multiple scattering processes can in a full quantum
mechanical treatment give rise to effects not found in a
semi-classical study.

\section{Discussion and conclusion}

Hybrid semiconductor-superconductor structures offer interesting
possibilities for investigating fundamental transport phenomena as
well as for potential applications. We have studied the possibility of
side-gate modulating the critical current in a mesoscopic
superconductor-2DEG-superconductor Josephson junction.  Side-gate
modulation offers a new alternative to devices based on field-effects,
non-equilibrium effects, and optical effects. In the case of side-gate
modulation the side-gate is used in tuning the transmission of the
2DEG to resonance.

Our calculations are based on a numerical treatment of the
transmission properties of the 2DEG region. Using the Landauer formula
\cite{landauer1957,landauer1970,fisher1981} and a similar formula for
the Josephson current \cite{beenakker1991b} we have then calculated
the essential transport properties: the normal state conductance
$G_{\rm N}$, the critical current $I_{\rm c}$, and the $I_{\rm
  c}R_{\rm N}$ product. The conductance in the superconducting state
can unfortunately not be expressed directly in terms of the
transmission eigenvalues and is complicated due to the presence of
multiple Andreev reflections (see however Ref. \cite{johansson1999}
for a possible solution). The considered geometry is comparable to the
situation in recent experiments \cite{bastian2000} even though some
simplifications have been made. A quantitative comparison would
require {\it i)} that the different Fermi properties of the
superconductor and the 2DEG are taken into account
\cite{mortensen1999a}, {\it ii)} more lattice points ($M_{\rm L}\gg
N$) in order to account correctly for the case of multi-mode leads
($N\sim 16$ in Ref. \cite{bastian2000}), and {\it iii)} that the width
of the 2DEG should be much larger than the separation $W'$ of the two
leads.  However, our model studies demonstrate how the side-gate
modulation can be used in controlling both the normal state
conductance and the critical current.

Our results are in qualitative agreement with the recent experimental
findings \cite{bastian2000} and confirm the possibility of modulating
the transport properties by a side-gate. Studies of the local density
of states confirm the existence of ``non-local'' modes and indicate
that more simple trajectory models can not fully account for the
detailed electron and hole propagation -- rather a full quantum
mechanical treatment is necessary.

\vspace{5mm}
\noindent 
{\it Acknowledgements} --- We would like to thank M. Brandbyge, A.-P.
Jauho, K. Flensberg, and H. Takayanagi for useful discussions.

\end{document}